\documentstyle[11pt,psfig,cite]{article}
\pssilent
\let\oldsubsec=\subsection\let\oldbfs=\bfseries
\def\subsection#1{\def\bfseries{\it}\oldsubsec{#1}
\let\bfseries=\oldbfs}

\def\beq{\begin{equation}}
\def\eeq{\end{equation}}
\def\r{{\bf r}}
\def\a{\alpha}\def\b{\beta}\def\e{\epsilon}\def\l{\lambda}
\def\rmin{r_{\it min}}
\def\w#1#2{w_{(#1,#2)}}
\def\fkm{Falicov--Kimball model}
\def\etal{{\it et al.}}

\begin{document}
\title{Repulsive particles on a two-dimensional lattice}
\author{G.~I.~Watson\\
  Rutherford Appleton Laboratory\\
  Oxfordshire OX11 0QX, UK}
\date{9 September 1996}
\maketitle

\begin{abstract}
The problem of finding the minimum-energy configuration of particles
on a lattice, subject to a generic short-ranged repulsive interaction,
is studied analytically.  The study is relevant to charge ordered states
of interacting fermions, as described by the spinless Falicov--Kimball
model.  For a range of particle density including the half-filled case,
it is shown that the minimum-energy states coincide with the large--$U$
neutral ground state ionic configurations of the Falicov--Kimball model,
thus providing a characterization of the latter as ``most homogeneous''
ionic arrangements.  These obey hierarchical rules, leading to a sequence
of phases described by the Farey tree.  For lower densities, a new family
of minimum-energy configurations is found, having the novel property that
they are aperiodic even when the particle density is a rational number.
In some cases there occurs local phase separation, resulting in an
inherent sensitivity of the ground state to the detailed form of the
interaction potential.
\end{abstract}

\section{Introduction}
The problem studied in this paper is the search for ground states of
a two-dimensional repulsive lattice gas at zero temperature, defined
as follows.  Suppose a collection of classical particles, which will be
called ``ions'', is allowed to occupy the sites of an infinite square
lattice, such that no two ions occupy the same site.  With each spatial
arrangement of ions is associated an energy density,
\beq {\cal E} = N^{-1}\sum_{i<j}V(\r_i-\r_j), \label{et}\eeq
where $\r_i$ is the position of the $i$th ion, $N$ is the number of
sites, and $V$ is some repulsive interparticle pair potential.  What is
the arrangement of ions which minimizes ${\cal E}$?

Specific forms of the interaction potential $V$ will be discussed below.
Of interest is the case where $V$ is short-ranged, in the sense that it
falls off rapidly with distance, so the contribution to the energy from
a given ion is dominated by a few near neighbours.  Nevertheless, the
long-range part of the interaction will be important also, in breaking
the degeneracy of configurations with identical short-range correlations.

Our work is motivated by two questions.  The first arises in the study
of the \fkm\ (Falicov and Kimball 1969).  This is a prototype model of
many-body correlations in an interacting fermion system, in which an
electron gas of density $\rho_e$ interacts with a collection of heavy
classical particles (``ions'') of density $\rho_i$ through an on-site
interaction of strength $U$.  It has several interesting interpretations,
including as a model of valence fluctuations in rare earth and transition
metal oxides, as a version of the Hubbard model, and as a model of
binary alloys (for background discussion, see Kennedy and Lieb 1986,
Lieb 1986, J\c{e}drzejewski, Lach and \L{}y\.zwa 1989a,b, Kennedy 1994,
Watson and Lema\'nski 1995, Gruber \etal~1995, Gruber and Macris 1996).
Let us focus on the particular case of the neutral model ($\rho_e=\rho_i$)
in the large--$U$ limit, and on the question of determining the ground
state ionic configurations, which are those for which the electrons'
energy is a minimum, at fixed density.  For the one-dimensional \fkm,
this problem has been solved: the ground states for rational densities
are the {\it most homogeneous} ionic configurations.  Lemberger (1992)
has provided both a precise definition of this property, and a proof
that it is satisfied for the large--$U$ neutral ground states.  Here,
we investigate the question of whether a similar homogeneity property can
be defined which characterizes the large--$U$ neutral ground states for
the {\it two-dimensional} model.  We approach the question via classical
energetics, since it is clear that at least some of the model's properties
can be understood as effects of repulsive effective interionic forces
(Watson and Lema\'nski 1995).

Thus, our problem captures one aspect of charge ordered states on a
two-dimensional lattice: the energetics of static classical interacting
particles.  Despite the idealized interparticle potential and the neglect
of kinetic energy and quantum effects, features emerge which appear to
be typical of two-dimensional charge ordering.  We find, for example,
phases in which charges line up to form slanted stripes, which are
also found in the \fkm\ (Kennedy 1994, Watson and Lema\'nski 1995).
Similar charge (and spin) stripe structures are observed in a family of
high-temperature superconductors (Tranquada \etal~1995), in its nickelate
(Tranquada \etal~1994) and manganate (Sternlieb \etal~1996) analogues,
and in other compounds (Chen and Cheong 1996).  In those cases the
mechanisms involved are very different from the model considered here,
but may also, in some cases, be dominated by static energetics (Emery
and Kivelson 1993).  Charge stripes are also found, for example, in
theories of two-dimensional interacting electrons in a magnetic field
(Koulakov, Fogler and Shklovskii 1996, Moessner and Chalker 1996).

The second issue we wish to address is of a more general nature.
The problem under study is one in which there is a competition between
two length scales.  One length scale is the lattice spacing, and the
other is the average interparticle distance.  The one-dimensional
version of the problem has been studied by Hubbard (1978) and others
(Pokrovsky and Uimin 1978, Burkov 1983, Aubry 1983a), in the context of
a model of quasi-one-dimensional conductors, and in connection with the
Frenkel--Kontorova model (Burkov, Pokrovsky and Uimin 1982).  In that
case, the repulsive forces encourage the ions to spread out evenly with
a neighbour distance $1/\rho$, where $\rho$ is the ion density, but this
distance may be incompatible with the constraint that ions occupy sites of
the background lattice.  The exact solution obtained by Hubbard for the
case of decreasing, convex potentials, has features typical of systems
with competing length scales, and in general of systems with modulated
phases resulting from frustrated interactions (Bak 1982, Aubry 1983b,
Heine 1987, Yeomans 1988, Lovesey, Watson and Westhead 1991).  When the
ion density is a rational number, the ground state ion configuration is
periodic, and may be constructed hierarchically using a simple branching
rule displaying the structure of a Farey tree (Aubry 1983a, Levitov
1991, Gruber, Ueltschi and J\c{e}drzejewski 1994).  The dependence
of ion density on chemical potential is a complete devil's staircase,
a general consequence of the convex form of the potential, which may
be understood using a formulation in terms of interacting domain walls
(Fisher and Szpilka 1987, Bassler, Sasaki and Griffiths 1991).

The two-dimensional generalization studied here displays a more dramatic
kind of lattice mismatch.  The natural ionic configuration in the
absence of a background lattice is a triangular lattice, with neighbour
distance $(2/\sqrt3\rho)^{1/2}$.  With a square background lattice,
the ions are always hindered from adopting their natural arrangement,
for any density: the ions would like to form a triangular lattice, but
are forced to occupy the sites of a square lattice.  Thus, increasing the
dimension to two introduces the possibility of a geometrical mismatch, in
addition to the simple length scale mismatch studied for one-dimensional
systems.  (In three dimensions, the problem is harder still, as then the
minimization of repulsive energy is nontrivial even in the absence of a
background lattice; see the remarks below concerning the relation with
sphere packing, a famous unsolved problem in three-dimensional geometry
(Croft, Falconer and Guy 1991, Conway and Sloane 1993)).

The problem of repulsive particles on a two-dimensional lattice is related
to other topics of recent interest.  As an example, we mention experiments
(Baert \etal~1995) investigating flux lines in a superconductor, in
the presence of an array of artificial antidots which pin the flux
lines to the sites of a square lattice.  Although the forces between
flux lines are not of the short-ranged type studied here, the kind of
lattice mismatch which occurs is the same.  In consequence, the pinning
force, derived from critical current measurements, shows peaks at certain
rational matching fields.  (We shall return to this example in Sec.~4.)
Hierarchical structures have been found previously in models of flux
line lattices (Levitov 1991).  Finally, we mention lattice gas models
of monolayers of adsorbed atoms on a lattice substrate (Pokrovsky
and Uimin 1978, Bak \etal~1979, Villain 1980), and observed in-plane
ordering of intercalant layers in graphite (Suzuki and Suematsu 1983,
Zabel and Chow 1986).

\section{Hierarchical structures in the \fkm}
This section presents a short summary of some aspects of the \fkm.

The Hamiltonian of the \fkm, in its usual spinless form, and in the
canonical ensemble (fixed $\rho_e$ and $\rho_i$), is
\beq
H = -\sum_{\langle ij\rangle}a_i^\dagger a_j-U\sum_iw_in_i, \label{fkm}
\eeq
where $a_i^\dagger$ is the creation operator for a spinless ``electron''
on site $i$ of a $d$-dimensional lattice, and $w_i$ is a classical
variable taking the value 1 if site $i$ is occupied by an ``ion'' and 0
otherwise.  Hopping is restricted to sites $i$ and $j$ which are nearest
neighbours, and $n_i=a_i^\dagger a_i$ is the electron occupation of site
$i$.  The Hamiltonian does not contain a kinetic term for the ions, so the
ions do not move.  However, the ground state is constructed by choosing
the particular configuration of ions (i.e.~the values of $w_i$) for which
the total electronic energy computed from $H$ is a minimum.  We take
$U>0$, so the electron--ion interaction is attractive.  This is not a
restriction, since the attractive and repulsive models are in one-to-one
correspondence with each other, via a particle--hole transformation.

The \fkm\ has been studied using a variety of techniques, for lattices in
one dimension and higher, and many interesting features have been found
(for a review of exact results, see Gruber and Macris 1996; for numerical
studies, see Watson and Lema\'nski 1995, Gajek, J\c{e}drzejewski and
Lema\'nski 1996, and references therein).  Here we focus on the large--$U$
neutral ground states.  It is in the limit of strong electron--ion
coupling that the behaviour of the model is simplest.  The phase
diagram in the grand canonical ensemble contains only segregated phases
(phase-separated mixtures of the vacuum and the phase in which every site
is occupied) and a family of neutral configurations ($\rho_e=\rho_i$).
As mentioned above, the neutral large--$U$ ground states of the
one-dimensional model, for rational densities, are fully characterized
by the property of being the most homogeneous (Lemberger 1992).

Qualitatively, the homogeneity property means that the ions would like
to be as far apart as possible, i.e.~that the effective forces between
them are repulsive.  The physical origin of this effective force may
be understood by imagining that each ion traps exactly one electron
in a bound state to form an atom, and that the interaction between
atoms is repulsive because the Pauli principle inhibits overlap of the
electrons' wavefunctions.  Analytical calculations of the effective
ion--ion interaction potential have been made (Gruber, Lebowitz and
Macris 1993a,b, Gruber, Ueltschi and J\c{e}drzejewski 1994, Watson and
Lema\'nski 1995, Micheletti, Harris and Yeomans 1996), in the main for
the one-dimensional model.  Gruber and co-workers (Gruber, Lebowitz and
Macris 1993a,b, Gruber, Ueltschi and J\c{e}drzejewski 1994) showed that in
one dimension the total energy is given to leading order in $t/U$ by a sum
of two-body potentials for neighbouring ions, and discussed qualitatively
three-body and higher potentials.  Micheletti, Harris and Yeomans (1996)
calculated the general $n$-body interaction to leading order.  The crucial
feature is that these potentials are exponentially decreasing convex
functions of the separation of the outermost ions, and it follows from
general arguments (Fisher and Szpilka 1987, Bassler, Sasaki and Griffiths
1991) that the ground state phase diagram exhibits a complete devil's
staircase, in which the ion configurations obey hierarchical branching
rules.  An alternative, if less rigorous, argument is to imagine that
convex repulsive forces exist between all pairs of ions, in which case
the work of Hubbard (1978) leads to the same conclusion.  Numerical work
on the one-dimensional \fkm\ is in perfect agreement with these ideas,
and it is fair to say that a coherent understanding has emerged.

\begin{figure}
\vskip6truemm
\centerline{\psfig{file=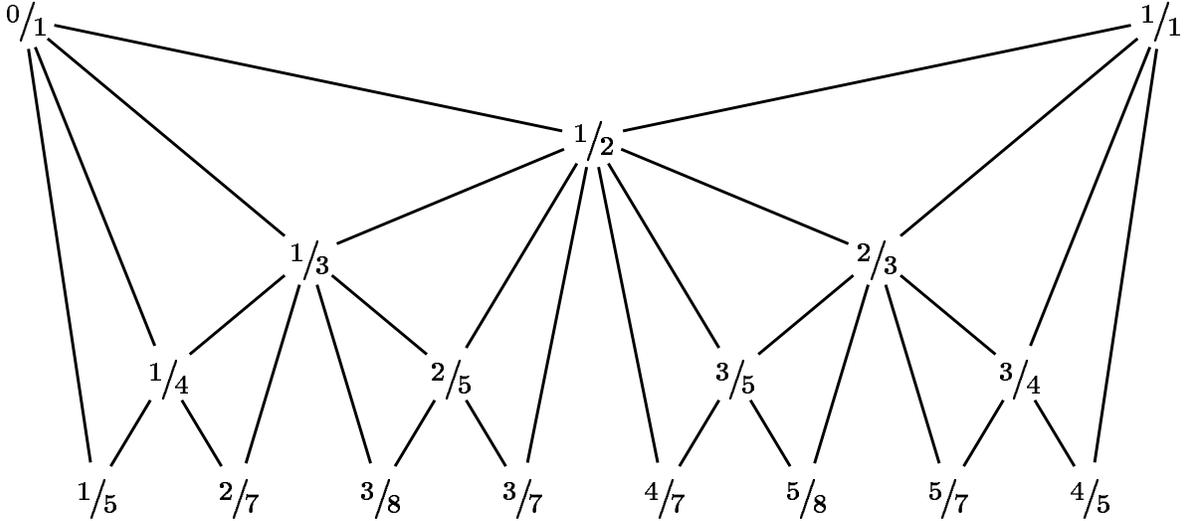,width=6.2truein,angle=90}}
\vskip-3truemm
\caption{\small The first five generations of the Farey tree,
according to which periodic large--$U$ neutral ground states of the
\fkm\ are constructed.  The ground state for a given rational density is
obtained by concatenating the unit cells of its two parents; for example,
density 3/8 is obtained by combining 1/3 and 2/5.}
\end{figure}

Let us describe the hierarchical construction of the most homogeneous
configurations in one dimension.  It is based on the Farey tree
(Aubry 1983a), part of which is displayed in Fig.~1.  The tree extends
downwards to infinity, with the rule that each pair of adjacent fractions
$p_1/q_1$ and $p_2/q_2$ is assigned a descendant $(p_1+p_2)/(q_1+q_2)$.
Every rational fraction between 0 and 1 occurs somewhere in the tree.
The recursive rule for constructing the ion configuration corresponding
to a given rational density $p/q$ is then that it has period $q$, with
a unit cell obtained by concatenating the unit cells of its parent
fractions in the Farey tree (the lower density parent on the left).
Thus, with the $\rho=0$ and $\rho=1$ states having unit cells (0) and
(1) respectively, $\rho=1/2$ corresponds to (01), $\rho=1/3$ to (001),
$\rho=2/5$ to (00101), and so on.  The construction given here can
be expressed in several equivalent forms: as a recursive criterion
for equal spacing of gaps between ions (Lemberger 1992), in terms
of the continued fraction expansion of the density (Burkov 1983),
in terms of the solutions of a Diophantine equation (Hubbard 1978,
Freericks and Falicov 1990), and as a ``circle sequence'' (Luck 1989).
Although the rigorous analysis does not extend to irrational values
of the density, one expects from continuity that the ground states
are given by a similar construction.  These are then quasiperiodic,
i.e.~one-dimensional quasicrystals.  The Fourier spectrum of such a
configuration has singularities on a dense set of points, namely the
frequency module generated by two noncrystallographic primitive vectors.

Let us now turn to the two-dimensional model, whose behaviour is less well
understood.  Published work to date include several preliminary studies
(J\c{e}drzejewski, Lach and \L{}y\.zwa 1989a,b, Gruber \etal~1990), a
collection of general analytical properties (Gruber, J\c{e}drzejewski
and Lemberger 1992), exact analyses of large--$U$ neutral equilibrium
states for certain densities, on square and triangular lattices, in
the presence of a magnetic field, and at zero and nonzero temperature
(Gruber, J\c{e}drzejewski and Lemberger 1992, Kennedy 1994, Gruber
\etal~1995, Messager and Miracle-Sole 1996), and a numerical calculation
(Watson and Lema\'nski 1995) of restricted phase diagrams in the grand
canonical ensemble.  The phase diagrams in two dimensions are more complex
than in one dimension, but there are some common features, especially
for large $U$ where the neutral phases dominate.  Kennedy (1994), in a
rigorous perturbative analysis for large $U$, proved that the neutral
ground states for densities $\rho=1/3$, 1/4 and 1/5 are those given in
Fig.~2, and provided a partial characterization of the ground states for
the range $1/4<\rho<1/2$.  Each ground state in this range consists of
(not necessarily equally spaced) stripes of occupied sites, slanted
at an angle which is a piecewise constant function of the density.
For $\rho=1/2$, the ground state is the ``checkerboard'' configuration
(Kennedy and Lieb 1986, Lieb 1986) given in the figure.  Watson and
Lema\'nski (1995), on the basis of numerical calculations, suggested
that such slanted stripe phases occur over a wider density range, and
that the structure perpendicular to the stripes obeys hierarchical rules
based on the Farey tree.

\begin{figure}
\vskip6truemm
\centerline{\psfig{file=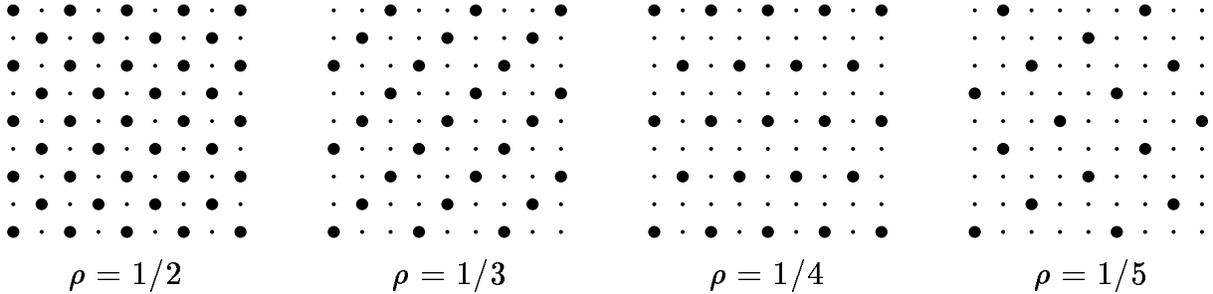,width=6.4truein,angle=90}}
\vskip-3truemm
\caption{\small Neutral ground state ionic configurations of the
two-dimensional large--$U$ \fkm, for densities 1/2, 1/3, 1/4 and 1/5.
Large dots: occupied sites; small dots: vacant sites.  These are also
ground states of the repulsive ion problem, with respect to the ``greedy
potential'' (see Sec.~4).}
\end{figure}

Thus, these conjectured ground states follow a hierarchical construction
identical to that for the one-dimensional model.  However, as pointed
out by Kennedy (1994), such configurations do not appear to be the most
homogeneous, even though the effective interactions are entirely repulsive
for large $U$ (Watson and Lema\'nski 1995).  One purpose of the present
work is to show that, at least for a range of densities, these ground
states {\it are} in fact the most homogeneous configurations, according to
a very reasonable criterion: they minimize the repulsive energy, ${\cal
E}$, with $V$ representing a ``generic'' short-ranged repulsion.  Thus,
as in one dimension, the physics is dominated by the tendency for ions
to spread as far apart as possible, and the results are not sensitive
to the precise form of the effective repulsion.  Having demonstrated
that ${\cal E}$ has the same ground states as the \fkm\ for the density
range $1/4\le\rho\le1/2$, in which exact results and reliable numerics
are available, we can use ${\cal E}$ to explore the possibilities for
other densities.

\section{The interparticle potential}
In the notation using site occupation variables $w_i$, the energy density
is
\beq {\cal E} = (1/2N)\sum_{ij}V_{i-j}w_iw_j = (1/2)\sum_jV_jg_j, \eeq
where the second equality writes it in terms of the pair correlation
function,
\beq g_i = (1/N)\sum_jw_jw_{i+j}. \eeq
We note that ${\cal E}$ satisfies a duality property: under
a particle--hole transformation, $w'_j=1-w_j$, the pair correlation
function becomes
\beq g'_i = 1-2\rho+g_i. \label{dual}\eeq
Working in the canonical ensemble, where the density is constant, the
pair correlation functions of dual configurations differ by a constant,
and hence so do the energies.  The conclusion is that the ground states
for densities $\rho$ and $1-\rho$ are duals of one another.  Consequently,
only densities less than 1/2 need be considered.

Qualitative features of the effective interionic potential for the
two-dimensional \fkm\ have been discussed previously (Watson and
Lema\'nski 1995) and are similar to the one-dimensional case (Gruber,
Lebowitz and Macris 1993a,b, Gruber, Ueltschi and J\c{e}drzejewski
1994, Micheletti, Harris and Yeomans 1996).  The effective potential
has a two-body part, whose long-range asymptotic form for large $U$
(Watson 1995) is proportional to $CR\exp(-R/\lambda)$, where $\lambda$
is a $U$-dependent ``atomic'' length, $R$ is the ``Manhattan distance'',
$R=m+n$, between ions at sites $(0,0)$ and $(m,n)$, and $C$ is the
combinatorial factor $(m+n)!/m!n!$, giving the number of walks from
one site to the other.  There are also multi-body potentials, evident,
for example, in calculated large--$U$ effective Hamiltonians (Gruber,
J\c{e}drzejewski and Lemberger 1992, Kennedy 1994, Gruber \etal~1995,
Messager and Miracle-Sole 1996) which include only short-range terms.

Here we are interested in properties which do not depend on any specific
numerical form of the interaction potential.  Let us then neglect
three-body and higher order forces, and attempt to abstract the idea of
a short-ranged two-body potential to the extreme of pure combinatorics.
We introduce our ``greedy potential'' as follows.  Rather than assigning
a numerical value to the energy of a configuration, the greedy potential
is a rule for deciding which of two configurations is energetically
preferred.  The rule is to compare the number of ions per unit area which
have neighbours at distance $r=1$, the lower number being preferred.
If the $r=1$ correlations of the two configurations are the same, one
compares the density of neighbours at $r=\sqrt2$.  If this does not break
the tie, one continues to distances $2$, $\sqrt5$, $\sqrt8$, and so on.
In other words, with the greedy potential the configurations are ordered
hierarchically with respect to radial correlations.  It follows from
(\ref{dual}) that the greedy potential obeys the duality property,
so that ground states at densities $\rho$ and $1-\rho$ are related by
a particle--hole transformation.

We emphasize that there is no rigorous basis for using the greedy
potential in the context of the \fkm.  The greedy potential depends
only on radial distance, whereas the correct pair potential is a strong
function of the Manhattan distance, in the sense that the potential
at distance $R+1$ is weaker than that at distance $R$ by a power of
$U$.  Furthermore, the true effective Hamiltonian includes multi-body
interactions entering at lower powers of $1/U$ than the longer range parts
of the two-body term.  Nevertheless, as follows from the work of Lemberger
(1992), for the {\it one-dimensional} \fkm\ the neutral large--$U$ ground
states coincide precisely with the ground states of the greedy potential.
Thus, in one dimension, the pair potential overwhelms other contributions,
and an interesting question is whether this occurs also in two dimensions.

\section{Ground states}
In the following sections we solve completely the problem of finding
the ground state ion configurations with the greedy potential, for the
density ranges $1/6\le\rho\le1/5$ and $1/4\le\rho\le1/2$, and we provide a
fairly complete characterization of the ground states for $1/5<\rho<1/4$.
To begin, we show that for a few special values of the density which
best match the background lattice, the ground states may be deduced
immediately from elementary considerations.

For any configuration, define $\rmin$ as the shortest distance between
any pair of ions.  Consider the set of ion configurations which have
$\rmin\ge\sqrt2$, i.e.~which do not contain any pairs of ions at nearest
neighbour distance, $r=1$.  We ask, what is the maximum ion density for
a configuration in this set?  Since the density is the inverse of the
average area of the Voronoi cells of the configuration, one may try to
minimize the Voronoi cell area of a single ion, subject to the constraint
that all other ions are further than $r=1$ away from the central ion and
from each other.  The optimum configuration is clearly that pictured in
Fig.~3(a), which has a Voronoi cell area of 2.  It happens that there
exists a configuration in the infinite square lattice for which every ion
has a local arrangement like this, namely the checkerboard configuration
in Fig.~2.  We conclude that if $\rmin\ge\sqrt2$ then $\rho\le1/2$, and
that this minimum is achieved only for the checkerboard configuration.
An equivalent statement is that if $\rho=1/2$ then $\rmin\le1$ for any
configuration except the checkerboard.  It follows that the checkerboard
is the ground state for density 1/2.

\begin{figure}
\vskip6truemm
\centerline{\psfig{file=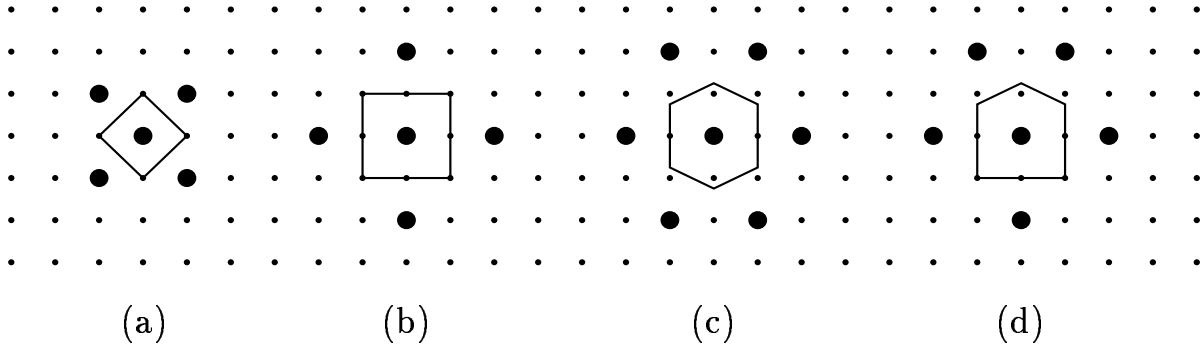,width=6.4truein,angle=90}}
\vskip-3truemm
\caption{\small Local ionic arrangements minimizing the area of
the Voronoi polygon (solid lines), subject to constraints on $\rmin$,
the distance of closest approach: (a) $\rmin\ge\protect\sqrt2$; (b)--(d)
$\rmin\ge2$.}
\end{figure}

A similar argument can be applied to configurations with the constraint
$\rmin\ge2$.  The smallest Voronoi cell area is 4, and up to rotations and
reflections there are three local configurations achieving the minimum,
illustrated in Fig.~3(b), (c) and (d).  The argument must now be extended
to take account of the multiplicity of $r=2$ correlations, but the logic
is essentially the same as above and leads to the conclusion that the
density $1/4$ ground state configuration is identical to that shown
in Fig.~2.

Continuing the same argument to lower densities is straightforward,
except that it becomes increasingly difficult to prove rigorously that a
particular set of local configurations have the minimum Voronoi cell area.
Using elementary inequalities (Watson 1995), this task can be reduced to
the laborious checking of a finite set of possibilities, which can then
be carried out by hand or by computer.  The results are as follows: for
density 1/5, the ground state is the configuration pictured in Fig.~2,
while for densities 1/8, 1/9, 1/10, 1/13, 1/15 and 1/18, the ground state
configurations are those shown in Fig.~4.  We have carried this analysis
through for densities as low as 1/90 (corresponding to $\rmin\ge10$);
the optimum configurations have the appearance of distorted triangular
lattices.  The only case found not to be amenable to this analysis is
$\rmin\ge\sqrt{20}$, for which the minimum area Voronoi cell is a pentagon
(area $\approx 19.97$) which fails to tile the plane.  The smallest
Voronoi cell which tiles the plane has area 20, so it appears likely
that the highest packing density with $\rmin\ge\sqrt{20}$ on an infinite
lattice is 1/20, but this has not been proved.

\begin{figure}
\vskip6truemm
\centerline{\psfig{file=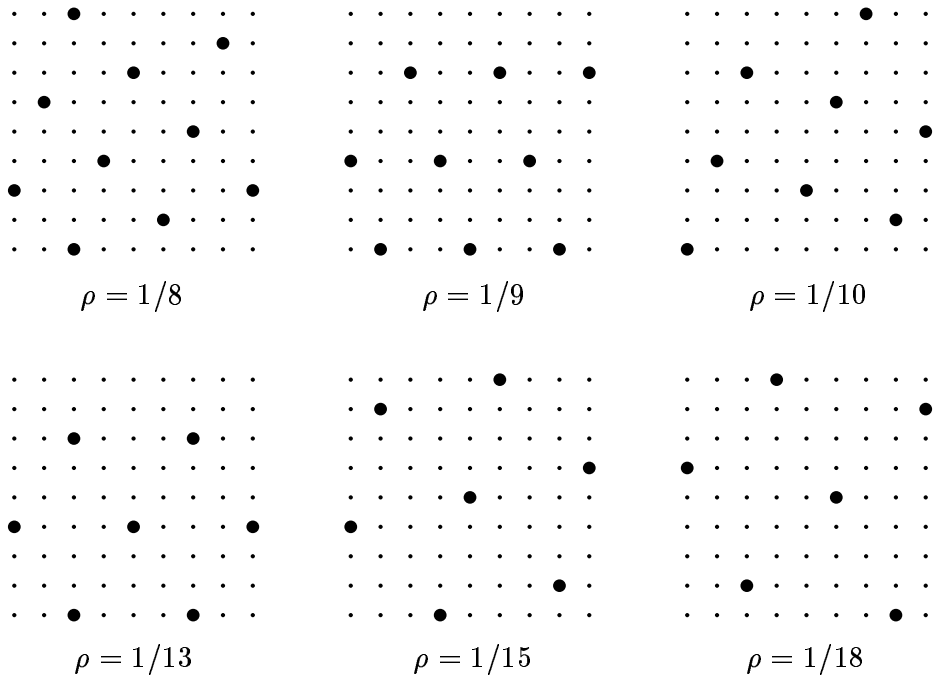,width=5truein,angle=90}}
\vskip-3truemm
\caption{\small Ion configurations shown to be ground states with
respect to the ``greedy potential'', for special matching densities $\rho$.}
\end{figure}

The crucial step in the above argument is the maximizing of the density
subject to a constraint relating to the distance of closest approach of
two ions.  It is interesting that the repulsive ion problem is, in this
respect, closely related to disc packing (Croft, Falconer and Guy 1991,
Conway and Sloane 1993).

The densities for which this simple disc packing argument yields a
rigorous determination of the ground state are the magic values 1/2, 1/4,
1/5, 1/8, and so on.  These may be regarded as densities of best fit for
which repulsive particles are ``most comfortable'' on a square lattice.
We note, in passing, that best fit configurations on a square lattice
have been considered previously by Baert \etal~(1995), who argued that
these determine the positions of anomalous peaks in the pinning force
measured in experiments on flux lines in a superconductor with a periodic
array of pinning centres.  However, the authors stated that the matching
configurations are tilted square lattices, which appears questionable,
assuming that the dominant factor is the mutual repulsion of flux lines.
Indeed, it is hard to imagine the tilted square lattices for $\rho=1/4$
or 1/8 being lower in energy than those in Figs.~2 and 4 for any
reasonable repulsive potential.  We would rather suggest that that
the flux line configurations corresponding to the rational matching
peaks are those in Figs.~2 and 4.  On the other hand, our suggestion
is not consistent with the claim that a peak is found at $\rho=1/16$.
Neither suggestion explains the absence of peaks at 1/9, 1/10 and 1/13
in the experimental data.

\subsection{Density $1/4\le\rho\le1/2$}
We now determine rigorously the ground states of the repulsive ion problem
(with the greedy potential) for a range of densities.  The technique used
is one previously applied to the \fkm\ (Gruber, J\c{e}drzejewski and
Lemberger 1992, Kennedy 1994, Gruber \etal~1995), based on decomposing
a Hamiltonian into contributions from $3\times3$ blocks.  In each case,
the property of minimizing the Hamiltonian does not fix a unique state,
but reduces the search for the ground state to a subset of possible
configurations.  Then additional reasoning is used to determine the
ground state uniquely.

First, we require an appropriate Hamiltonian.  Let us introduce
\beq H_r = (1/N)\sum_{\langle ij\rangle=r}w_iw_j, \eeq
where the sum is over all bonds of length $r$.  Thus, $H_r$ counts the
density of pairs of ions at distance $r$.  Consider the density range
$1/3\le\rho\le1/2$.  It follows from the argument of the previous section
that $\rmin\ge\sqrt2$, and the bound $\rmin\le(2/\sqrt3\rho)^{1/2}$
therefore implies $\rmin=\sqrt2$.  Hence, by construction, each
ground state for the greedy potential in $1/3\le\rho\le1/2$ minimizes
$H_{\sqrt2}$.  However, this condition is not sufficient to determine the
ground states uniquely, as may be seen from the examples given in Fig.~5.
The configurations pictured both have density 2/5, and they also have
identical $r=\sqrt2$ correlations.  It is necessary to go to $d=2$
to reveal that (a) is preferred.  Thus, we introduce a total Hamiltonian
\beq H = 4H_{\sqrt2} + 6\e H_2, \eeq
which includes an $r=2$ term weighted by a small coefficient $\e$,
to break the tie between configurations degenerate at $r=\sqrt2$.
(The coefficients 4 and 6 are included for later convenience.)  We shall
use this Hamiltonian to determine the ground state in the density range
$1/4\le\rho\le1/2$.

\begin{figure}
\vskip6truemm
\centerline{\psfig{file=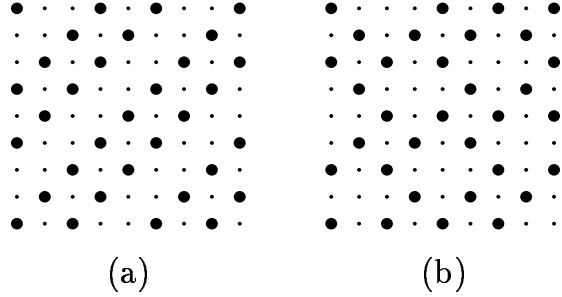,width=3truein,angle=90}}
\vskip-3truemm
\caption{\small Two configurations with density 2/5, having identical
correlations up range $r=\protect\sqrt2$.  Configuration (a) is preferred
when $r=2$ correlations are taken into account.}
\end{figure}

Let us now define a set of local functions, referring only to sites
within a $3\times3$ block.  Labelling the sites within a block by local
coordinates $(i,j)$ with $0\le i,j\le2$, we write
\begin{eqnarray}
h_1 &=& \w11\\
h_2 &=& \w00+\w20+\w22+\w02\\
h_3 &=& \w01+\w10+\w21+\w12\\
h_4 &=& \w01\w10+\w10\w21+\w21\w12+\w12\w01\\
h_5 &=& \w00\w11+\w20\w11+\w22\w11+\w02\w11\\
h_6 &=& \w00\w20+\w01\w21+\w02\w22+\w00\w02+\w10\w12+\w20\w22.
\end{eqnarray}
These satisfy
\beq 4\sum_Bh_1 = \sum_Bh_2 = \sum_Bh_3 = 4N\rho \eeq
\beq \sum_Bh_4 = \sum_Bh_5 = 2NH_{\sqrt2} \eeq
\beq \sum_Bh_6 = 3NH_2, \eeq
where $\sum_B$ denotes the sum over all blocks.

The approach is now as follows.  One constructs a block Hamiltonian
$h$ which is a linear combination of $h_1$ to $h_6$, in such a way
that $N^{-1}\sum_Bh$ differs from $H$ only by a term proportional
to the density.  Since we are working in the canonical ensemble, the
difference is a constant, and minimizing $N^{-1}\sum_Bh$ is equivalent
to minimizing $H$.  One then writes out all possible configurations of
ions in a single $3\times3$ block: there are 20 possibilities, up to
rotations and reflections, listed by Kennedy (1994).  One determines
a subset of block configurations such that $h$ is minimized on that
subset, for all sufficiently small $\e$.  Then, provided there exists at
least one configuration of the required density on the infinite lattice
which is made up entirely of blocks in that subset, it follows that the
ground state for the greedy potential is made up entirely of blocks in
the subset.

Clearly, the trick is to find a suitable block Hamiltonian.  Here we
shall just state a choice that works.  For $1/3\le\rho\le1/2$, let
\beq h = -(4+4\e)h_1-(1+4\e)h_2-(2+4\e)h_3+h_4+h_5+2\e h_6, \eeq
and for $1/4\le\rho\le1/3$, let
\beq h = -(4-6\e)h_1-(1-\e)h_2-(2-2\e)h_3+(1+\e)h_4+(1-\e)h_5+2\e h_6. \eeq
The remaining analysis is identical to that used by Kennedy (1994) for
the \fkm: in his notation, the former $h$ is found to be minimized for
block configurations 12, 13, 17, 19 and 20, while the latter is found
to be minimized for configurations 2, 5, 6, 11, 12 and 13.  These are
precisely the same block configurations which determine the ground states
of the \fkm\ in the same density ranges, and the same conclusion follows
immediately (Kennedy 1994).  For $1/3\le\rho\le1/2$, the ground states
consist of fully occupied stripes of slope 1 separated by unoccupied
sites, i.e.~the ground states are invariant under translations by
$(1,1)$ or $(1,-1)$.  (Examples of such configurations are the first two
configurations in Fig.~2, and Fig.~5(a).)  For $1/4\le\rho\le1/3$, the
ground states consist of stripes of slope $1/2$, i.e.~are invariant under
translation by one of the vectors $(2,1)$, $(1,2)$, $(2,-1)$ or $(-1,2)$.

The stripe property does not determine the configuration uniquely.
However, it does reduce the problem to a one-dimensional one, since the
only remaining freedom is the arrangement of stripes in the perpendicular
direction.  When restricted to the set of configurations satisfying
the stripe property, the greedy potential leads to a purely repulsive
interaction between stripes.  In fact, this interaction is described
by the one-dimensional form of the greedy potential, and it follows
from Hubbard's (1978) result that the stripe arrangement is the ``most
homogeneous'' one, constructed according to the Farey tree as described
in Sec.~2.  We have therefore arrived at a complete description of the
ground states, in this density range, of the repulsive ion problem with
the greedy potential.  A recursive construction of the ground states
for rational densities can be made, as follows.  The $\rho=1/2$ state
has primitive vectors A and C, where $A=(-1,1)$ and $C=(1,1)$.  It is
convenient to think of it as being constructed by placing a stripe
generated by C, then a second identical stripe separated from the
first by A, a third stripe separated from the second by A, and so on.
The structure transverse to the stripes is then $\ldots AAA\ldots$,
denoted (A).  Similarly, the $\rho=1/3$ state is (B), where $B=(-2,1)$.
Their descendant, the $\rho=2/5$ state, is then (AB), which is the
configuration in Fig.~5(a).  The $\rho=3/7$ ground state is (ABB),
and so on.  The ground states in $1/4\le\rho\le1/3$ are generated in
the same manner by $A=(-1,1)$, $B=(0,2)$ and $C=(2,1)$.

The ground states of our repulsive ion problem and the neutral
large--$U$ ground states of the \fkm\ have the same stripe structure
in this density range.  This is not sufficient to deduce that the two
sets of ground states are identical.  However, the numerical evidence
that the Farey tree rules are obeyed for the ground states of the \fkm\
is most persuasive (Watson and Lema\'nski 1995).  It seems reasonable,
therefore, to conjecture that the repulsive ion problem does in fact
have the same ground states as the \fkm, i.e.~that the greedy potential
captures the dominant physics in the neutral large--$U$ limit, for this
range of densities.  This result suggests that there is a well-defined
sense in which the large--$U$ neutral ground states of the \fkm\ could
be regarded as the ``most homogeneous'' configurations.

As Kennedy (1994) has noted, the stripe structure does not accord
with plausible intuitive ideas about what homogeneity might mean in
two dimensions.  As an example, he imagines forming a configuration
with density slightly less than 1/2 by starting with the checkerboard
and removing ions in such a way that these vacancies are as far apart as
possible; the configuration in Fig.~5(b) is of this type.  This example
raises an interesting point.  If a finite number of vacancies is
introduced into the checkerboard configuration they cannot destroy
the long-range order, and it is clear that they will repel each other,
in order to maximize the number of neighbouring ions whose repulsive
energy is reduced.  If the number of vacancies is increased beyond a
certain point, presumably of order $N^{1/2}$, the vacancies suddenly
come together to form a domain wall, and the full long-range order of
the checkerboard is destroyed.

\subsection{Density $1/5\le\rho\le1/4$: General structure}
For densities less than 1/4, we have $\rmin\ge2$, i.e.~no two ions may be
closer than $r=2$.  This implies that the nine of the twenty $3\times3$
block configurations do not occur in the ground state.  Defining the
block Hamiltonian
\beq h = -3h_1-h_2-2h_3+h_6, \eeq
we find that the block configurations 2, 5, 8 and 11 (in Kennedy's
notation) minimize $h$, amongst the remaining eleven possible blocks.
Since $N^{-1}\sum_Bh$ is a multiple of $H_2$, apart from a term
proportional to the density, and since the ground state lies in the set of
configurations which minimize $H_2$, it follows that the ground state is
composed entirely of blocks of type 2, 5, 8 and 11.  The same property was
shown by Kennedy to hold for the \fkm.  It is not sufficient to determine
the ground state uniquely, but it does place strong constraints on it.

Each block centred on an occupied site is of type 2.  The block one
site to the right of a type 2 block must be type 5, type 8 (in one of
two orientations) or type 11.  Considering all possibilities for adjacent
blocks, one finds as in Fig.~6 that the ions to the right of any given ion
(circled) must be arranged as in (a), (b) (with two possible orientations)
or (c).

\begin{figure}
\vskip6truemm
\centerline{\psfig{file=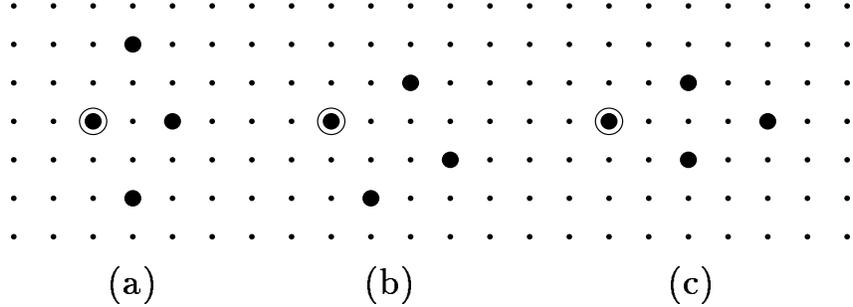,width=4.5truein,angle=90}}
\vskip-3truemm
\caption{\small Possible configurations of ions to the immediate
right of any given ion (circled), for a ground state configuration in
the density range $1/5\le\rho\le1/4$.}
\end{figure}

Therefore, we may join ions by lines in such a way that each ion is a
vertex of a quadrilateral lying to its right.  That quadrilateral is
either a ``diamond'' in one of two orientations, as in (a) or (c), or a
square in one of two orientations, as in (b).  The same applies to the
three other orthogonal directions from any central ion.  Hence each ion
is the common vertex of four such quadrilaterals.  The ion configuration
has a square lattice ``skeleton'', in the sense that the lattice formed
by the ions, with edges drawn to form quadrilaterals as described,
is topologically equivalent to a square lattice.  Equivalently, the
configuration could be described as a tiling of the plane by two kinds of
tiles, the diamond and the square.  Configurations made up only of blocks
of type 2, 5, 8 and 11 are in one-to-one correspondence with such tilings.

Next, we consider the structure of the tilings.  One can construct a
tiling by beginning with a tiling of squares only, and then introducing
one or more ``kinks'', where a kink is a line of diamonds, running along
one of the two possible orthogonal directions.  Two orthogonal kinks
may intersect, as illustrated in Fig.~7(a).  Clearly, any arrangement of
kinks is a valid tiling, and conversely, it can be shown that any tiling
corresponds to an arrangement of kinks.  For any tile configuration,
the ion density does not depend on the arrangement of kinks, but only
on the kink {\it densities} $\a$ and $\b$ in the two directions:
$\rho=1/(5-\a-\b+2\a\b)$.  The problem is to find the best kink
configuration, subject to the constraint of constant density.

\begin{figure}
\vskip6truemm
\centerline{\psfig{file=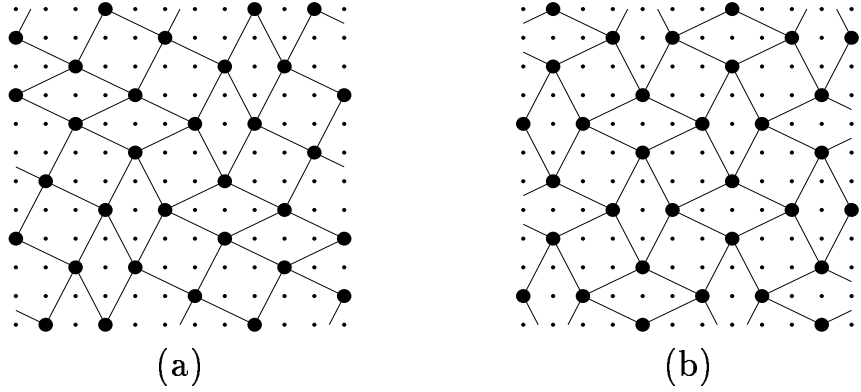,width=4.5truein,angle=90}}
\vskip-3truemm
\caption{\small (a) An ion configuration corresponding to a plane
tiling in which two orthogonal kinks intersect. (b) A configuration with
density 2/9, constructed from a tiling in which the kink density is 1/2
in both directions.  The lines are a guide to the eye, showing the
corresponding tile arrangement.}
\end{figure}

The definition of a kink is such that a transformation which turns a
kink into an antikink (the absence of a kink) leaves all configurations
invariant, up to rotation.  Since this transforms the kink densities as
$\a\to1-\a$ and $\b\to1-\b$, we need consider only $\a\le1/2$.

The argument so far depends only on the information as to which $3\times3$
block configurations may occur in the ground state, and applies equally
in the case of the \fkm.  To take further the analysis for the greedy
potential, we include longer range correlations by considering possible
configurations of four tiles meeting at a vertex.  There are four cases,
illustrated in Fig.~8, A--D.  If a configuration is such that a fraction
$\l_A$ of its ions correspond to type A vertices, a fraction $\l_B$
to type B vertices, and so on, with $\sum\l_i=1$, then its density is
\beq \rho = (5\l_A+9\l_B/2+9\l_C/2+4\l_D)^{-1}. \eeq
The weights of correlations at particular distances, or, equivalently,
values of the radial distribution function $g(r)$, are related to the
fractions $\l_i$ of the four types of vertex.  We find
\begin{eqnarray}
g(2) &=& (\l_B+\l_C)/2+\l_D = 5-1/\rho\\
g(\sqrt5) &=& 2\\
g(\sqrt8) &=& g(3) = 0\\
g(\sqrt{10}) &=& 2\l_A+\l_B+\l_C = 2(1/\rho-4)\\
g(\sqrt{13}) &=& \l_C/2+\l_D = 5-1/\rho-\l_B/2.\label{ctt}
\end{eqnarray}
All except the last of these is independent of the $\l$ coefficients.
Hence, it is necessary to go at least to $r=\sqrt{13}$ correlations to
break the degeneracy of tile configurations.  From (\ref{ctt}), {\it
the ground state belongs to the set of configurations for which $\l_B$,
the fraction of type B vertices, is maximum.}

\begin{figure}
\vskip6truemm
\centerline{\psfig{file=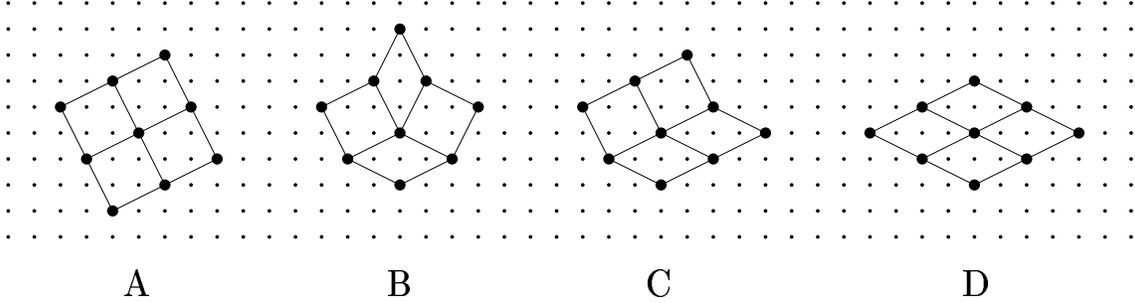,width=6truein,angle=90}}
\vskip-3truemm
\caption{\small The four possible types of vertex which can occur
in a configuration corresponding to a tiling of the plane by diamonds
and squares.}
\end{figure}

Type B sites occur only where kink-antikink neighbour pairs intersect
pairs running in the orthogonal direction.  Their density is maximized
for kink densities less than 1/2 by requiring all kinks to be isolated,
and for kink densities greater than 1/2 by requiring all antikinks to
be isolated.  If $\a$ and $\b$ are both less than or equal to 1/2, this
gives a B site density $\l_B=4\a\b$, which must now be minimized subject
to the constraint of fixed density.  The optimum is easily shown to be
\beq \a = \b = [1-(9-2/\rho)^{1/2}]/2, \eeq
so the kink densities are the same in both directions.  This case
corresponds to $\rho\le2/9$.  If $\a\le1/2$ and $\b\ge1/2$, we have
$\l_B=\a(1-\b)$, and the optimum now has the density of kinks in the
$\a$ direction equal to the density of antikinks in the $\b$ direction,
i.e.~$\b=1-\a$.  The two cases may be written as
\beq \rho = {2\over9\pm(1-2\a)^2}, \label{yhw}\eeq
where the sign is plus for $\rho<2/9$ and minus for $\rho>2/9$.

Special cases of interest are $\a=0$ and $\a=1/2$.  The former
yields the $\rho=1/4$ and $\rho=1/5$ ground states of Fig.~2, which
correspond to diamond and square tilings, respectively.  The latter
gives a $\rho=2/9$ ground state with alternating kinks and antikinks,
which is the configuration shown in Fig.~7(b).

We have shown that the ground state consists of a distribution of isolated
kinks or antikinks in a tilted square lattice background, and we have
determined the kink densities.  The precise arrangements of the kinks will
be considered in the next subsection.  First, let us note one interesting
consequence of the relationship (\ref{yhw}) between kink density and ion
density.  If the ground state ion configuration is periodic, then the kink
configuration is periodic and the kink density must be a rational number.
It follows that the ion density is of the form $\rho=2/[9\pm(p/q)^2]$,
with $p$ and $q$ integers such that $0\le p/q\le1$.  For all ion densities
other than this set of special values, the ground state configuration is
certainly not periodic, because the kink density is an irrational number.
The ion densities for which an aperiodic ground state is found includes
``most'' rational values.  This is a somewhat surprising result: in
previous investigations of related models, it has been found that the
ground state is periodic whenever it is possible for it to be periodic,
i.e.~whenever the density is a rational number.

\subsection{Density $1/5\le\rho\le1/4$: Kink configurations}
Having determined the kink densities, it is necessary to study the
kink--kink interaction to find the arrangement of kinks in the ground
state.  We already know that the interactions are such as to rule out
nearest-neighbour kinks or antikinks, i.e.~they are repulsive at the
shortest range.

It is possible to study the kink interactions systematically beginning
with short range terms.  For example, given a configuration containing
the segment $000101$, where 1 denotes a kink and 0 denotes an antikink,
one may examine the change in energy on moving the kink by one step
to $001001$.  Such a move involves shifting a single row of ions by
(-1,1), or a vector equivalent under rotation.  The energy change depends
on the arrangement of kinks in the perpendicular direction.  In this
case, however, it can be shown that, regardless of the perpendicular
structure, the energy is always lowered by an amount of magnitude at
least $V(\sqrt{34})$.  Since a move of the form $10010001$ to $10001001$
always gives an energy change at a range greater than $\sqrt{34}$,
it follows that no ground state contains both 101 and 10001 segments.
Hence, for kink densities $1/3\le\a\le1/2$, the configuration is made up
entirely of kinks spaced 2 or 3 units apart, while for $\a\le1/2$, the
ground state does not contain any kinks spaced 2 units apart.  In other
words, kinks are purely repulsive at range 2, implying that the ground
states for $1/3\le\a\le1/2$ have {\it homogeneous} kink configurations,
in the terminology of Lemberger (1992).  However, the kink configurations
may or may not be the {\it most} homogeneous, depending on longer range
interactions.

If kink interactions were purely repulsive in all cases, the ground
states would indeed be those with the most homogeneous kink arrangements.
That this is not true can be seen in the example of Fig.~9.  Here we see a
kink configuration $00001001$ along an axis (2,1).  If kink interactions
were repulsive at range 3, the energy could be lowered by moving the
central kink one unit to the left, yielding the configuration $00010001$.
This move corresponds to shifting a row of ions by (-1,1), to the sites
shown in the figure as open circles.  The resulting changes in radial
correlations all cancel out to range $\sqrt{58}$.  The lowest order
change in energy comes at range $\sqrt{61}$, from the pair of sites
joined by an arrow in the figure, and is {\it positive}.  Hence the
range 3 kink interaction is attractive, and this conclusion holds for
any kink structure in the perpendicular direction with $0<\a\le1/3$.

\begin{figure}
\vskip6truemm
\centerline{\psfig{file=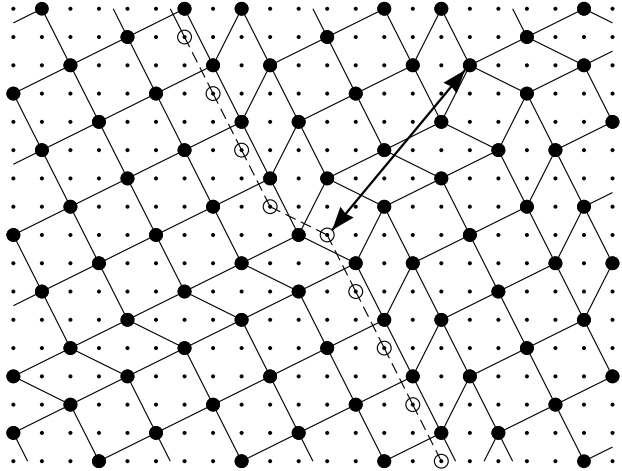,width=3.3truein,angle=90}}
\vskip-3truemm
\caption{\small Configuration illustrating the attractive kink interactions
at range 3.  Moving the central kink one unit to the left, corresponding
to shifting a row of sites by (-1,1) to the open circles, incurs an energy
cost at range $\protect\sqrt{61}$, represented by the arrow.}
\end{figure}

The consequence of such an attractive kink force is dramatic.  Since,
with the greedy potential, a correlation at range $\sqrt{61}$ can never be
outweighed by longer range correlations, no matter how large their weight,
it is always energetically favourable for two kinks to come together to
a distance of 3 units.  The result is phase separation: for $\a<1/3$, all
the kinks will clump together in a region in which all neighbouring kinks
are 3 units apart, and the rest of the lattice will be free of kinks.
This occurs in both directions, so the result is four distinct regions:
one with density 1/5 (kink free), two regions of density 3/14 (kink
free in one direction but not the other), and one region of density 9/41
(intersecting kinks with spacing 3 in both directions).

We have argued above that the greedy potential provides a sensible
definition of most homogeneous configurations in two dimensions.  The
appearance of phase separation in the ground state kink configurations
reveals one of the limitations of this idea.  It would be perverse
to regard configurations with macroscopic phase separation as the most
homogeneous.  Nevertheless, the occurrence of attractive kink interactions
is an informative result.  Here, phase separation is a consequence of
the singular nature of the greedy potential, coming from the fact that
$V(r_1)$ is considered infinitely larger than $V(r_2)$ if $r_2>r_1$.
For a real interaction potential, where $V(r_2)$ is a finite, though
perhaps small, fraction of $V(r_1)$, long-range forces would prevent
macroscopic phase separation.  Instead, the result would be {\it local}
phase separation, yielding a state which appears inhomogeneous (phase
separated) on short length scales but is macroscopically homogeneous.
This is an example of frustrated phase separation (Emery and Kivelson
1993).  The magnitude of the intermediate length scale is sensitively
dependent on the relative magnitude of long-range and short-range
components of the potential.  No matter how small the long-range part,
it will eventually be magnified to a macroscopic scale as the system
tends toward phase separation.

Thus, the occurrence of phase separation in our model with the greedy
potential suggests that in two dimensions, and in this density range,
{\it the ground state is inherently sensitive to the details of the
interaction potential.} It also suggests that a useful and rigorous
definition of a unique ``most homogeneous'' configuration in two
dimensions is likely to be elusive.

The exact kink structure of the ground state is the outcome of the
competition between long-range forces and the tendency towards phase
separation.  In addition, there is the complicating factor that there are
two arrays of kinks intersecting one another, with kink interactions in
one direction coupled to the kink positions in the other.  The problem of
finding the ground states for a given model potential appears difficult,
and of limited value.  Let us merely note a few qualitative properties.
The appearance of attractive kink interactions leads to an incomplete
devil's staircase, with a tendency to `lock in' to a rational kink
density, such as 1/3 in the example described above.  For nearby
densities, there is local phase separation, stabilized by a repulsive
interaction on an intermediate length scale.  On this scale, there are
domains of kink-rich and kink-free structure.  One may then look at the
effective forces between domains, and apply the same reasoning to them
as for the kinks: if domain interactions are attractive at some range,
there is a tendency to phase separation and longer-range interactions
come into play.

It may be that for some density range the kinks (or domains) are purely
repulsive, independently of the kink arrangement in the orthogonal
direction.  In that case, we may conclude that they adopt a ``most
homogeneous'' configuration in each direction separately, and that
in a restricted interval the devil's staircase is complete.  The kink
arrangement is periodic if the kink (or domain) density is a rational
number, and quasiperiodic if it is irrational.  In the latter case,
the ground state is a two-dimensional quasicrystal.

\subsection{Density $1/6\le\rho\le1/5$}
The analysis for lower densities is very similar to the preceding cases.
For $\rho\le1/5$, we have $\rmin\ge\sqrt5$, and we may eliminate from
consideration all block configurations except types 1, 2, 3, 4, 8 and 9
(using the labelling of Kennedy 1994).  We define a new block function,
\begin{eqnarray}
h_7 &=& \w00\w12+\w00\w21+\w20\w01+\w20\w12 \nonumber\\
     && +\w02\w10+\w02\w21+\w22\w01+\w22\w10,
\end{eqnarray}
which counts $r=\sqrt5$ correlations, and a block Hamiltonian
\beq h = -2h_1-h_2-2h_3+h_7, \eeq
for which $N^{-1}\sum_Bh$ equals $2H_{\sqrt5}$ plus a term proportional
to the density.  Since $h$ is minimized for block types 2, 3, 8 and 9,
the ground state is composed entirely of those block configurations.

By considering the possible blocks in the immediate neighbourhood of a
given ion, one may show that the local environment of an ion reduces to
just a few possibilities.  The conclusion is similar to the case in the
previous section: the ground state configuration has a square lattice
``skeleton'', and by drawing appropriate bonds between neighbours can
be constructed from a tiling of the plane by polygons.  The tiles in
this case are a square, a parallelogram and a ``kite'', shown in Fig.~10.

\begin{figure}
\vskip6truemm
\centerline{\psfig{file=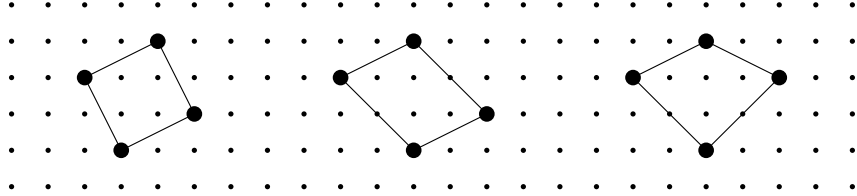,width=4.5truein,angle=90}}
\vskip-3truemm
\caption{\small The square, parallelogram and kite tiles, from which
the ground state ion configurations in the density range
$1/6\le\rho\le1/5$ are constructed.}
\end{figure}

Amongst all possible tilings, we must now select the ground state
by considering correlations at $r>\sqrt5$.  It is easy to see that
correlations up to $r=\sqrt{13}$ are the same for all tilings for a
given ion density, but that $g(4)$ equals the density of kite tiles
and is therefore minimized by choosing this density to be zero.  Thus,
the ground state corresponds to a tiling of the plane by squares and
parallelograms only.  In such a tiling, the arrangement of tiles at an
ion, corresponding to a vertex at which four tiles meet, is limited
to the four types shown in Fig.~11.  Once again, we rely on longer
range correlations to reduce the number of possibilities: for all
tilings of fixed density, correlations are equal for $r<\sqrt{18}$,
and $g(\sqrt{18})$ is nonzero only when vertices of type D are present.
The ground state tiling is made up only of type A, B and C vertices.

\begin{figure}
\vskip6truemm
\centerline{\psfig{file=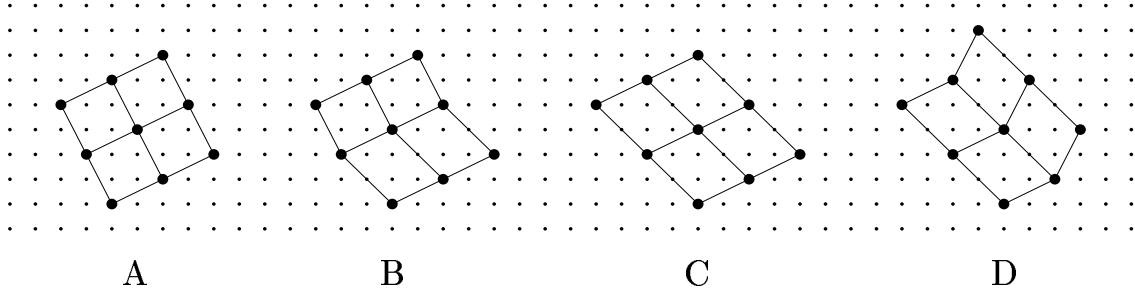,width=6truein,angle=90}}
\vskip-3truemm
\caption{\small The four possible types of vertices at which four tiles
meet, in a tiling of the plane by squares and parallelograms.}
\end{figure}

We now have sufficient information to specify the ground state uniquely.
A tiling using only these three types of vertex corresponds to a {\it
slanted stripe} configuration, of a similar kind to those found in
Sec.~4.1 to be ground states for $1/4\le\rho\le1/2$.  In this case
the slope of the stripes is 1/2.  The structure of the ground state
configurations perpendicular to the stripes is obtained, as before,
by considering the stripe interaction, which turns out once again to
be purely repulsive.  From Hubbard's (1978) result for one-dimensional
systems, the conclusion is that the spacing of the stripes follows the
``most homogeneous'' arrangement, constructed using the hierarchical
rules described previously.  In the notation of Sec.~4.1, the primitive
vectors of the ground states are $A=(-1,2)$, $B=(-2,2)$ and $C=(2,1)$.

We note, in passing, that the ground state configuration for ion density
1/6 (which is a tiling using only parallelograms in configuration C), does
not coincide with the configuration conjectured by Watson and Lema\'nski
(1995) to be the $\rho=1/6$ neutral large--$U$ ground state of the \fkm.
The latter is a tiling using parallelograms in configuration D, and is
therefore higher in energy according to the greedy potential.

The analysis could readily be continued to lower densities.  However, for
$\rho<1/6$, the ground states involve block configuration 1 (the empty
block), and it is necessary to go to $4\times4$ blocks to obtain useful
information using the block Hamiltonian approach.  We shall stop here.

\section{Discussion}
In the previous section we have succeeded in determining rigorously the
ground states of our repulsive lattice gas model in the density ranges
$1/6\le\rho\le1/5$ and $1/4\le\rho\le1/2$.  In the range $1/5<\rho<1/4$ we
have found a characterization of the ground states in terms of maximally
intersecting kink configurations, and have discussed qualitatively
the consequences of kink interactions.  (By particle--hole duality,
the results apply also to the range $1/2\le\rho\le5/6$.)

The results are not rigorously applicable to the \fkm.  As described in
Sec.~3, our ``greedy potential'' has been chosen to mimic qualitative
aspects of the effective interionic potential in the large--$U$ limit
of the neutral \fkm, but the two potentials differ in their details.
Their two-body components do not have the same dependence on the
separation of the ions, and the greedy potential does not include
three-body and higher order forces.

Nevertheless, we have observed that the ground states of the two models
are the same in all cases ($\rho=1/2$, 1/3, 1/4, 1/5) where both are
known, and that the repulsive lattice gas ground states are consistent
with all known properties of the \fkm\ ground states.  These are, namely,
that the ground states for $1/4\le\rho\le1/2$ satisfy the stripe property
(Kennedy 1994; see Sec.~4.1).  They are also consistent with numerical
results (Watson and Lema\'nski 1995) suggesting that the arrangement of
the stripes follows a hierarchical pattern, according to the Farey tree.
The properties of the ground states found here for $1/5<\rho<1/4$
are not consistent with the configurations conjectured by Watson and
Lema\'nski (1995) to be ground states of the \fkm\ in the same range,
but this could be due to the limited set of trial configurations used
in their numerical work.

The agreement between the ground states of the two models provides a
characterization of the large--$U$ neutral ground states of the \fkm.
For a limited range of densities, it is possible to define homogeneity in
a natural way, such that the \fkm\ ground states are the most homogeneous
ion configurations on the square lattice.  Potentially, this provides
a generalization of a result known rigorously to hold for all densities
in the one-dimensional \fkm.

Of equal importance is the finding that the repulsive ion model does {\it
not} provide a useful definition of the most homogeneous configuration
for the density range $1/5<\rho<1/4$, because of the tendency for
phase separation.  It shows that the ground state configuration is
inherently sensitive to the precise form of the forces between the ions.
This suggests that it is unlikely to be possible to characterize the
corresponding large--$U$ neutral ground states of the \fkm\ as unique
``most homogeneous'' configurations.  Instead, there is a whole class
of configurations which could be described as homogeneous, a qualitative
discussion of which has been given in Sec.~4.3.

Thus, although there is not a full correspondence between ground
states found here and those in the \fkm, our approach is of value
in exploring the possibilities.  As we have seen, the possibilities
are rather interesting.  The stripe property, which is satisfied for
the greedy potential ground states for densities $1/4\le\rho\le1/2$,
breaks down for densities less than 1/4 (although stripe phases re-enter
below $\rho=1/5$).  In the range $1/5<\rho<1/4$, the ground states do
not have this essentially one-dimensional structure; instead, they
consist of arrays of maximally intersecting kinks, or domain walls.
Amongst the interesting properties of this novel phase is the fact that
for almost all densities, including rational numbers, the ground state
is not periodic.  The only possible exceptions are densities that can
be written $2/[9\pm(p/q)^2]$ for integer $p$ and $q$.

The repulsive lattice gas studied here is one of a family of models,
involving two competing length scales, that has been investigated
extensively in one dimension.  The ground states of the one-dimensional
model have properties which are characteristic of a wide range of
related systems: they are periodic for rational densities and follow a
construction according to hierarchical rules.  Our results suggest that
a similar generality is possessed by the two-dimensional model.  All the
ground states we have found can be described as arrays of linear domain
walls in an otherwise periodic configuration of ions.  In some cases
these domain walls align parallel with one another, and stripe phases
result, which are periodic when the density is rational.  In other cases,
it happens that an intersection of domain walls has negative energy,
in which case the ground state is a maximally intersecting pattern of
domain walls.  Here, the densities for which it is possible for the
ground state configuration to be periodic are a rational function of
the {\it square} of a rational number.

This picture is analogous to the situation in one dimension, where
the ground states can also be represented as configurations of
(zero-dimensional) domain boundaries.  The crucial feature of two
dimensions is that two domain walls may intersect, and the intersection
energy may be positive or negative, leading to two distinct kinds of
ground states.  It would be interesting to investigate the obvious
extension of this idea to three dimensions.

\section*{Acknowledgements}
I would like to thank Romuald Lema\'nski and Martin Long for valuable
discussions.

\section*{References}
\def\epl#1{{\sl Europhys.~Lett.}~{\bf #1}}
\def\jcp#1{{\sl J.~Chem.~Phys.}~{\bf #1}}
\def\jpa#1{{\sl J.~Phys.~A.: Math.~Gen.}~{\bf #1}}
\def\jpc#1{{\sl J.~Phys.~C.: Solid State Phys.}~{\bf #1}}
\def\jpcm#1{{\sl J.~Phys.: Condens.~Matter} {\bf #1}}
\def\jpsj#1{{\sl J.~Phys.~Soc.~Japan} {\bf #1}}
\def\jsp#1{{\sl J.~Stat.~Phys.}~{\bf #1}}
\def\pha#1{{\sl Physica} A {\bf #1}}
\def\phb#1{{\sl Physica} B {\bf #1}}
\def\phc#1{{\sl Physica} C {\bf #1}}
\def\pla#1{{\sl Phys.~Lett.}~A {\bf #1}}
\def\pr#1{{\sl Phys.~Rev.}~{\bf #1}}
\def\pra#1{{\sl Phys.~Rev.}~A {\bf #1}}
\def\prb#1{{\sl Phys.~Rev.}~B {\bf #1}}
\def\prl#1{{\sl Phys.~Rev.~Lett.}~{\bf #1}}
\def\rmp#1{{\sl Rev.~Mod.~Phys.}~{\bf #1}}
\def\zpb#1{{\sl Z.~Phys.}~B {\bf #1}}
\def\\{\par}
\everypar={\hangindent=1cm\hangafter=1}\frenchspacing\parindent0pt
S.~Aubry 1983a {\sl J.~Physique Lett.}~{\bf 44} L247\\
S.~Aubry 1983b \jpc{16} 2497\\
P.~Bak, D.~Mukamel, J.~Villain and K.~Wentowska 1979 \prb{19} 1610\\
P.~Bak 1982 {\sl Rep.~Prog.~Phys.}~{\bf 45} 587\\
M.~Baert, V.~V.~Metlushko, R.~Jonckheere, V.~V.~Moshchalkov and
  Y.~Bruynseraede 1995 \epl{29} 157\\
K.~E.~Bassler, K.~Sasaki and R.~B.~Griffiths 1991 \jsp{62} 45\\
S.~E.~Burkov, V.~L.~Pokrovsky and G.~Uimin 1982 \jpa{15} L645\\
S.~E.~Burkov 1983 {\sl J.~Physique Lett.}~{\bf 44} L179\\
C.~H.~Chen and S.-W.~Cheong 1996 \prl{76} 4042\\
H.~T.~Croft, K.~J.~Falconer and R.~K.~Guy 1991 {\sl Unsolved
  Problems in Geometry} (New York: Springer)\\
J.~H.~Conway and N.~J.~A.~Sloane 1993 {\sl Sphere Packings,
  Lattices and Groups,} 2nd ed.~(New York: Springer)\\
V.~J.~Emery and S.~A.~Kivelson 1993 \phc{209} 597\\
L.~M.~Falicov and J.~C.~Kimball 1969 \prl{22} 997\\
M.~E.~Fisher and A.~M.~Szpilka 1987 \prb{36} 5343\\
J.~K.~Freericks and L.~M.~Falicov 1990 \prb{41} 2163\\
Z.~Gajek, J.~J\c{e}drzejewski and R.~Lema\'nski 1996 \pha{223} 175\\
C.~Gruber, J.~Iwanski, J.~J\c{e}drzejewski and P.~Lemberger 1990 \prb{41}
  2198\\
C.~Gruber, J.~J\c{e}drzejewski and P.~Lemberger 1992 \jsp{66} 913\\
C.~Gruber, J.~L.~Lebowitz and N.~Macris 1993a \epl{21} 389\\
C.~Gruber, J.~L.~Lebowitz and N.~Macris 1993b \prb{48} 10783\\
C.~Gruber, D.~Ueltschi and J.~J\c{e}drzejewski 1994 \jsp{76} 125\\
C.~Gruber, N.~Macris, A.~Messager and D.~Ueltschi 1995 ``Ground states
  and flux configurations of the two-dimensional \fkm'', preprint\\
C.~Gruber and N.~Macris 1996 ``The \fkm: a review of exact results and
  extensions'', preprint\\
V.~Heine 1987, in {\sl Competing Interactions and Microstructures: Statics
  and Dynamics,} edited by R.~LeSar, A.~Bishop and R.~Heffner (Berlin:%
  Springer), p.~2\\
J.~Hubbard 1978 \prb{17} 494\\
J.~J\c{e}drzejewski, J.~Lach and R.~\L{}y\.zwa 1989a \pla{134} 319\\
J.~J\c{e}drzejewski, J.~Lach and R.~\L{}y\.zwa 1989b \pha{154} 529\\
T.~Kennedy and E.~H.~Lieb 1986a \pha{138} 320\\
T.~Kennedy 1994 {\sl Rev.~Math.~Phys.}~{\bf 6} 901\\
A.~A.~Koulakov, M.~M.~Fogler and B.~I.~Shklovskii 1996 \prl{76} 499\\
P.~Lemberger 1992 \jpa{25} 715\\
L.~S.~Levitov 1991 \prl{66} 224\\
E.~H.~Lieb 1986 \pha{140} 240\\
S.~W.~Lovesey, G.~I.~Watson and D.~R.~Westhead 1991
  {\sl Int.~J.~Mod.~Phys.}~B {\bf 5} 1313\\
J.~M.~Luck 1989 \prb{39} 5834\\
A.~Messager and S.~Miracle-Sole 1996 {\sl Rev.~Math.~Phys.}~{\bf 8} 271\\
C.~Micheletti, A.~B.~Harris and J.~M.~Yeomans 1996 ``A complete devil's
  staircase in the \fkm'', preprint OUTP-96-22S\\
R.~Moessner and J.~T.~Chalker 1996 ``Exact results for interacting
  electrons in high Landau levels'', preprint\\
V.~L.~Pokrovsky and G.~V.~Uimin 1978 \jpc{11} 3535\\
B.~J.~Sternlieb, J.~P.~Hill, U.~C.~Wildgruber, G.~M.~Luke, B.~Nachumi,
  Y.~Moritomo and Y.~Tokura 1996 \prl{76} 2169\\
M.~Suzuki and H.~Suematsu 1983 \jpsj{52} 2761\\
J.~M.~Tranquada, D.~J.~Buttrey, V.~Sachan and J.~E.~Lorenzo 1994 \prl{73}
  1003\\
J.~M.~Tranquada, B.~J.~Sternlieb, J.~D.~Axe, Y.~Nakamura and S.~Uchida 1995
  {\sl Nature} {\bf 375} 561\\
J.~Villain 1980, in {\sl Ordering in Strongly Fluctuating Condensed Matter
  Systems,} ed.~T.~Riste (New York: Plenum), p.~221\\
G.~I.~Watson and R.~Lema\'nski 1995 \jpcm{7} 9521\\
G.~I.~Watson 1995 unpublished\\
J.~Yeomans 1988, in {\sl Solid State Physics}, edited by H.~Ehrenreich and
  D.~Turnbull (San Diego: Academic), Vol.~41, p.~151\\
H.~Zabel and P.~C.~Chow 1986 {\sl Comments Cond.~Mat.~Phys.}~{\bf 12} 225\\
\end{document}